\documentclass[12pt,preprint]{aastex}

\newcommand{\axp}{1E\ 1841-045}
\newcommand{\gtap}{\mathrel{\hbox{\rlap{\lower.55ex \hbox {$\sim$}}
                   \kern-.3em \raise.4ex \hbox{$>$}}}}
\newcommand{\ltap}{\mathrel{\hbox{\rlap{\lower.55ex \hbox {$\sim$}}
                   \kern-.3em \raise.4ex \hbox{$<$}}}}

\slugcomment{}

\shorttitle{Detection of hard pulsed X-rays from \axp}
\shortauthors{Kuiper et al.}

\begin{document}


\title{Discovery of hard non-thermal pulsed X-ray emission\\ from the anomalous X-ray pulsar \axp}

\author{L. Kuiper\altaffilmark{1}, W. Hermsen\altaffilmark{1,2} and M. Mendez\altaffilmark{1,2}}
\email{L.M.Kuiper@sron.nl}
\altaffiltext{1}{SRON-National Institute for Space Research, Sorbonnelaan 2, 3584 CA, Utrecht, The Netherlands }
\altaffiltext{2}{Astronomical Institute ``Anton Pannekoek", University of Amsterdam, Kruislaan 403, 1098 SJ Amsterdam, The Netherlands}


\begin{abstract}
We report the discovery of non-thermal pulsed X-ray/soft $\gamma$-ray emission up to $\sim 150$ keV from the anomalous X-ray pulsar 
AXP \axp\ located near the centre of supernova remnant Kes 73 using RXTE PCA and HEXTE data. The morphology of the double-peaked 
pulse profile changes rapidly with energy from 2 keV up to $\sim 8$ keV, above which the pulse shape remains more or less stable. The 
pulsed spectrum is very hard, its shape above 10 keV can be described well by a power law with a photon index of $0.94\pm 0.16$. 
\axp\ is the first AXP for which such very-hard pulsed emission has been detected, which points to an origin in the magnetosphere of a 
magnetar. We have also derived the total emission spectrum from the Kes73/\axp\ complex for the $\sim$ 2-270 keV energy range using 
RXTE HEXTE and XMM-Newton PN data. A comparison of the total emission from the complex with the pulsed+DC emission from \axp\ as derived from 
Chandra ACIS CC-mode data \citep{morii03} leaves little room for emission from Kes 73 at energies near 7 keV or above. This suggests 
that the HEXTE spectrum above $\sim$ 10 keV, satisfactorily described by a power law with index $1.47 \pm 0.05$, is dominated by emission 
from \axp. In that case the pulsed fraction for energies above 10 keV would increase from about 25\%  near  10 keV to 100\%  near
100 keV. The origin of this DC-component extending up to $\sim$ 100 keV is probably magnetospheric and could be a manifestation of
pulsed emission which is ``on" for all phases.
\end{abstract}







\section{Introduction}

The quest whether anomalous X-ray pulsars (AXPs) and soft gamma-ray repeaters (SGRs)
are both manisfestations of isolated neutron stars with ultra-strong magnetic fields 
($\sim$ $10^{14}-10^{15}$ G), so-called ``magnetars" \citep{thompson96}, has recently been decided.
Notably, the first detection of SGR-like bursts from the direction of AXP 1E 1048.1-5937 \citep{gavriil02},
followed by the detection of SGR-like bursts from AXP 1E 2259+586 \citep{kaspi03} conclusively 
unified AXPs and SGRs \citep[see also][]{gavriil04}. Similarly, \citet{kulkarni03} showed that the originally 
classical SGR 0526-66 was recently found to behave like an AXP. This was predicted uniquely by the 
magnetar model \citep{thompson96}. For recent reviews on AXPs and SGRs see \citet{mereghetti02} 
and \citet{woods04}.

In this paper we present detailed timing and spectral characteristics of AXP \axp, located
in the centre of the supernova remnant (SNR) Kes 73 (G27.4+0.0), which is at a kinematic distance
between 6 and 7.5 kpc \citep{sanbonmatsu92}. The X-ray pulsations of the source were discovered with 
ASCA \citep{vasisht97} and a phase-connected timing solution was published by \citet{gotthelf02}, who 
analyzed observations with  the Rossi X-ray Timing Explorer (RXTE) spanning 2 years. A linear ephemeris appeared to
be consistent with the pulse periods measured over 15 years with Ginga, ASCA, RXTE and BeppoSAX.
The measured constant, long-term spin-down of \axp\ as well as the phase-connected timing with
``timing-noise-like" residuals supported a magnetar identification.

\citet{morii03} reported detailed results for  \axp\ from Chandra ACIS CC-mode observations.
They could for the first time discriminate the compact object from the surrounding SNR Kes 73.  Like other AXPs,
the phase integrated spectrum (pulsed and DC emission) was well fitted with a power law (photon index 2.0 $\pm$ 0.3)
plus blackbody model (kT$_{\hbox{\tiny \rm BB}}$= 0.44 $\pm$ 0.02 keV). The photon index is the flattest among AXPs. 
It should be noted, however, that a two-blackbody fit rendered a similarly good fit. They also reported
that the pulse profile is double-peaked, with the second pulse exhibiting a harder spectrum between 3 and 7 keV.

\citet{molkov04} published recently a source catalog of 28 sources detected by INTEGRAL/IBIS between 18 and 60 keV
in a survey of the Sagittarius Arm tangent region. The source in this catalog with the hardest spectrum was identified with  \axp\ 
and SNR Kes 73. Since AXPs are ``known" to be soft-spectrum sources, the SNR seemed at first glance the most likely
counterpart. This identification stimulated us to analyse archival RXTE PCA and HEXTE data to search for
a pulsed signature from \axp, particularly in the hard X-ray range covered by HEXTE. We also analyzed
archival XMM-Newton data in an attempt to unravel the contributions from Kes 73 and the pulsed and unpulsed emissions
from \axp, combining the informations from our RXTE PCA/HEXTE and XMM-Newton analyses, and the Chandra results 
on \axp\ published by \citet{morii03} .


\section{Instruments and observations}
In this study the results come mainly from the analysis of data from two of the three X-ray instruments aboard RXTE , the Proportional Counter Array 
(PCA; 2-60 keV) and the High Energy X-ray Timing Experiment (HEXTE; 15-250 keV). Both are non-imaging instruments.
The PCA \citep{jahoda96} consists of five collimated xenon proportional counter units (PCUs) with a total effective area of $\sim 6500$ cm$^2$
over a $\sim 1\degr$ (FWHM) field of view. Each PCU has a front Propane anti-coincidence layer and three Xenon layers which provide the
basic scientific data, and is sensitive to photons with energies in the range 2-60 keV. The energy resolution is about 18\% at 6 keV.

The HEXTE instrument \citep{rothschild98} consists of two independent detectors clusters, each containing four Na(Tl)/CsI(Na) scintillation
detectors. The HEXTE detectors are mechanically collimated to a $\sim 1\degr$ (FWHM) field of view and cover the 15-250 keV energy range
with an energy resolution of $\sim$ 15\% at 60 keV. The collecting area is 1400 cm$^2$ taking into account the loss of the spectral capabilities of
one of the detectors. The maximum time resolution of the tagged events is $7.6\mu$s. In its default operation mode the field of view of each
cluster is switched on and off source to provide instantaneous background measurements.
Due to the co-alignment of HEXTE and the PCA, they simultaneously observe the sources in their field of view. Table \ref{obs_table} lists
the publicly available RXTE observations used in this study\footnote{\axp\ is regularly monitored
by RXTE since February 1999}. All observations listed in Table \ref{obs_table} have been carried out with the instrumental pointing axis within 
$\sim 0\farcm 5$ from \axp. In the fourth column the PCU unit 2 screened exposure is given (see Sect. \ref{pca_timing}). A typical observation
run consists of several sub-observations spaced more or less uniformly between the start and end date of the observation.
In this study the total number of sub-observations amounts 45.

\begin{table}[t]
\caption{List of  RXTE observations used in this study of \axp.\label{obs_table}}
\begin{center}
\begin{tabular}{cccr}
\hline\hline
\textbf{Obs.} & \multicolumn{2}{c}{\textbf{Begin/End Date}}    & \textbf{Exp.$^{\dagger}$}\\
\textbf{id.}  & \multicolumn{2}{c}{\textbf{(dd/mm/yyyy)}}      & \textbf{(ks)}\\
\hline
40083            & 15-02-1999          & 23-02-2000           & 105.73 \\
50082            & 10-04-2000          & 08-03-2001           &  59.91 \\
60069            & 02-04-2001          & 26-01-2002           &  52.24 \\
70094            & 16-03-2002          & 27-01-2003           &  53.26 \\
\hline
\multicolumn{4}{l}{$^{\dagger}$PCU-2 exposure after screening} \\
\end{tabular}
\end{center}
\end{table}


\section{Timing analysis}

\subsection{Timing analysis of PCA data}
\label{pca_timing}
The PCA data from the observations listed in Table \ref{obs_table} have all been collected in {\em Goodxenon} or {\em GoodxenonwithPropane}
 mode allowing high-time-resolution ($0.9\mu$s) studies in 256 spectral channels. Because we are mainly interested in the medium/hard X-ray timing
 properties of \axp\ we ignored the events triggered in the Propane layers of each PCU. Furthermore, we used only data from the top xenon layers of each
 PCU in order to improve the signal-to-noise ratio. Because the number of active PCUs at any time was changing we treated the five PCUs
 constituting the PCA separately. For each PCU good time intervals have been determined by including only time periods when the PCU in question is on 
 and during which the pointing direction is within $0\fdg 05$ from the target, the elevation angle above Earth's horizon is greater than $5\degr$, a time delay 
 of 30 minutes since the peak of a South-Atlantic-Anomaly passage holds, and a low background level from contaminating electrons is observed. 
 These good time intervals have subsequently been applied in the screening process to the data streams from each of the PCUs (e.g. see Table 
 \ref{obs_table} for the resulting screened exposure of PCU-2 per observation run).

 \begin{figure}[t]
  \epsscale{0.5}
  \plotone{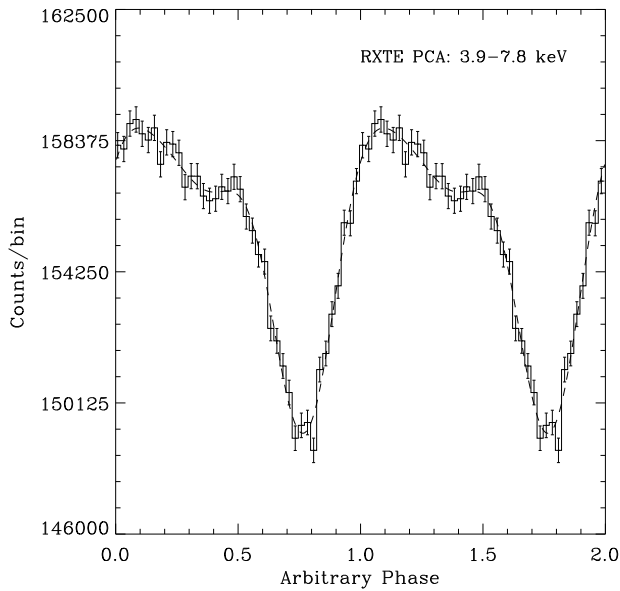}
  \caption{\label{pca_pulse_template}RXTE PCA pulse profile of \axp\ for energies in the range 3.9-7.8 keV combining data collected between
  February 1999 and February 2003 (see Table \ref{obs_table}). Two cycles are shown for clarity. The best fitting truncated Fourier series
  (3 harmonics) is superposed as a dashed line.}
\end{figure}

Next, for each sub-observation the arrival times of the selected events (for each PCU unit) have been converted to arrival times at the solar
 system barycenter (in TDB time scale) using the instantaneous spacecraft position and celestial position of \axp\  \citep[see][]{gotthelf02}.
 These barycentered arrival times have been folded with the phase connected timing solution given in \citet{gotthelf02} using only the first
 three frequency coefficients to obtain pulse phase distributions for selected energy windows. Combining now the phase distributions from the various PCUs
 the well-known 2-10 keV profile of \axp\  \citep[see][]{gotthelf02} could be recognized in each sub-observation, however, phase shifts between 
 the sub-observations made a direct combination impossible.
 Therefore we correlated the pulse phase distribution of each sub-observation with a chosen initial template and applied the measured phase shifts
 to obtain an aligned combination with much higher statistics. The correlation is then once repeated with, instead of the initially chosen template, the aligned
 combination from the first correlation to obtain the final summed profile \citep[see e.g.][for a similar iterative method applied for PSR B0540-69]{deplaa03}.
 The summed profile for energies between 3.9-7.8 keV deviates from uniformity at a $\sim 32\sigma$ level and is shown in Fig. \ref{pca_pulse_template}. 
 Its morphology mimics the profile shown by \citet{gotthelf02}. For the first time, however, pulsed emission has been detected up to $\sim 25$ keV -
 the deviation from uniformity is $9.5\sigma$ and $3.2\sigma$ in the 11.7-16.1 keV and 16.1-23.8 keV energy ranges, respectively - providing
 strong evidence for a non-thermal origin of the pulsed high-energy emission.
 This result motivated us to search also the higher-energy HEXTE data for a pulsar signal.

\begin{figure}[t]
  \epsscale{0.5}
  \plotone{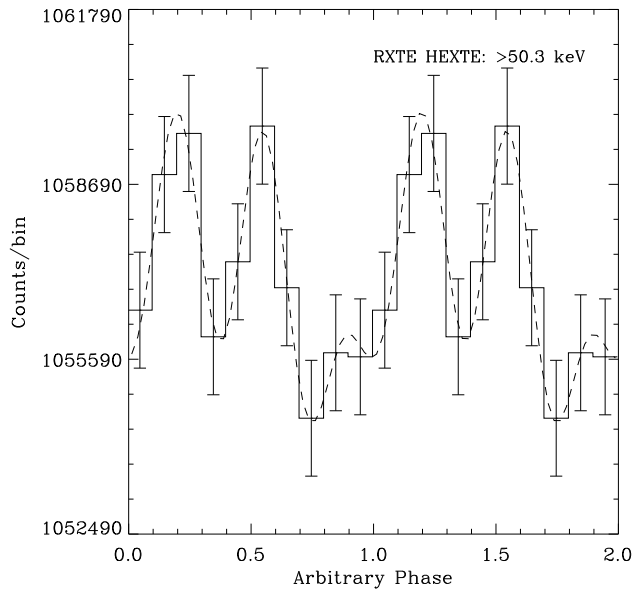}
  \caption{\label{hexte_pulse_prof_gt50} RXTE HEXTE pulse profile of \axp\ for energies greater than 50.3 keV demonstrating the existence
  of pulsed emission from this source at hard X-rays/soft $\gamma$-rays. The deviation from uniformity is about $4.1\sigma$. The dashed line is the
  best fitting truncated Fourier series (3 harmonics) like shown in Fig. 1. The structured profile and the minimum level 
  approximately coincide in phase with those in the 3.9-7.8 keV profile of Fig. \ref{pca_pulse_template}}.
\end{figure}
\begin{figure*}[t]
  \centerline{\includegraphics[height=16cm,width=10cm,angle=90]{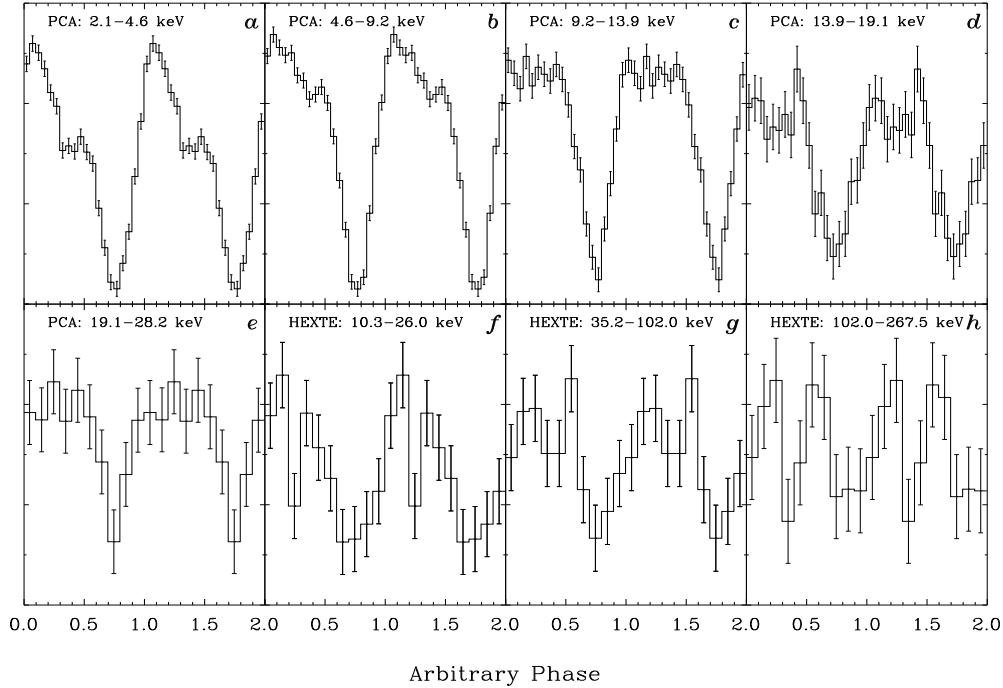}}
  \caption{\label{profile_collage}A RXTE PCA/HEXTE collage of pulse profiles of \axp\ for energies greater than 2.1 keV. Two cycles are shown
  for each panel. Panels {\em a-e}: RXTE PCA pulse profiles for energies between 2.1 and 28.2 keV. Panels {\em f-h}: RXTE HEXTE pulse
  profiles for the energy ranges 10.3-26 keV, 35.2-102 keV and 102-267.5 keV. The three HEXTE profiles deviate from uniformity at a 4.2, 3.1
  and 2.7$\sigma$ level, respectively. Notice the morphology change of the pulse profiles, particularly in the PCA profiles.}
\end{figure*}

\subsection{Timing analysis of HEXTE data}

HEXTE operated in its default rocking mode during the observations listed in Table \ref{obs_table} allowing the 
collection of real-time background data
from two independent positions $\pm 1\fdg 5$ to either side of the on-source position. For the timing analysis 
we selected only the on-source data.
Good time intervals have been determined using similar screening filters as used in the case of the PCA. The 
selected on-source HEXTE event times
have subsequently been barycentered and folded according to the ephemeris given in \citet{gotthelf02} using 
again only the first three frequency coefficients.
Applying the phase shifts as derived from the contemporaneous PCA measurements to the HEXTE phase distributions of each sub-observation 
we could obtain the HEXTE pulse phase distributions in 256 spectral channels for the combination of observations listed in Table \ref{obs_table}. 
The HEXTE integral ($>10.3$ keV) pulse profile deviates from uniformity at a $\sim 4.7\sigma$ level 
and shows strong similarities with the PCA profile for energies above $\sim 10$ keV. Most remarkably,
the pulse profile for energies greater than 50.3 keV reaches a significance of $\sim 4.1\sigma$, 
proving the existence of pulsed emission from \axp\ at 
hard X-rays/ soft $\gamma$-rays (see Fig. \ref{hexte_pulse_prof_gt50}).

\subsection{Energy dependence of the pulse profile morphology}

\begin{figure}[h]
  \epsscale{0.45}
  \plotone{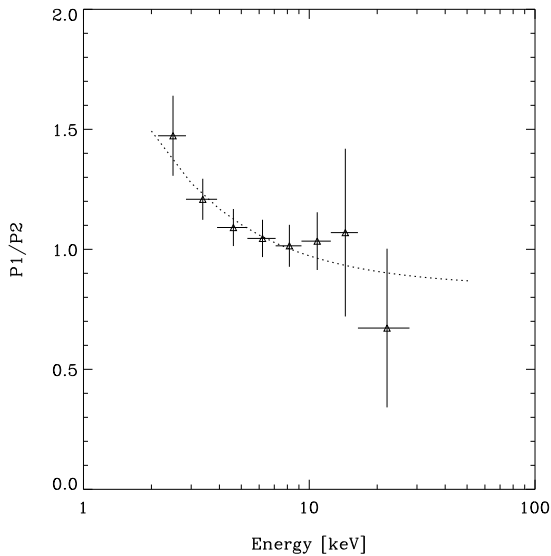}
  \caption{\label{p1_p2_ratio} The ratio of excess counts in phase range 0.8-1.25 (P1) and 0.25-0.7 (P2) for RXTE PCA
  data as a function of energy. The dotted line represents the best fit to a function of form $a+b/E$ with $(a,b)$ parameters to be 
  optimized. The improvement relative to a constant function $a$ is about $3\sigma$.}
\end{figure}

We have investigated the morphology of the pulse phase distributions as a function of
energy using both the RXTE PCA and HEXTE data. This kind of information could provide clues to the
origin and physics of the processes responsible for the pulsed high-energy radiation. 
In Fig. \ref{profile_collage} a collage 
of pulse-phase distributions is shown for different energy bands between $\sim 2$ and $\sim 270$ keV
\footnote{For HEXTE the 26-35.2 energy window is ignored in this collage because of a huge background line
near 30 keV due to the activation of iodine.}. 
A striking feature is that pulse minimum occurs in the phase range $[0.7,0.8]$ irrespective of the energy band.
Furthermore, the structured broad profile, measured with high statistics in the PCA data, seems to behave
as two components (``pulses") separated $\sim 0.4$ in phase, with the relative contributions varying with energy. 
The first ``pulse" near phase 0.1 is dominant over the second ``pulse" near phase 0.5 for energies
below $\sim 9$ keV. Above this energy the second pulse, however, has a similar intensity level as the first pulse. 
Quantitatively this behaviour is shown in Fig. \ref{p1_p2_ratio}, where the ratio of the number of excess counts in 
phase range 0.8-1.25 (P1) and 0.25-0.7 (P2) for RXTE PCA data is plotted as a function of energy (phase range 0.7-0.8 is chosen 
as a DC-reference window).
The $\chi^2$ of a fit to a function parameterized as $a+b/E$ improves by about $3\sigma$ (1 degree of freedom) relative
to the $\chi^2$ of a fit assuming a constant, i.e. no energy dependence of the ratio. Thus, we conclude that the morphology of the 
high-energy ($> 2$ keV) pulse profiles of \axp\ changes with energy.

\begin{figure}[t]
  \centerline{\includegraphics[height=9cm,width=11cm,angle=0]{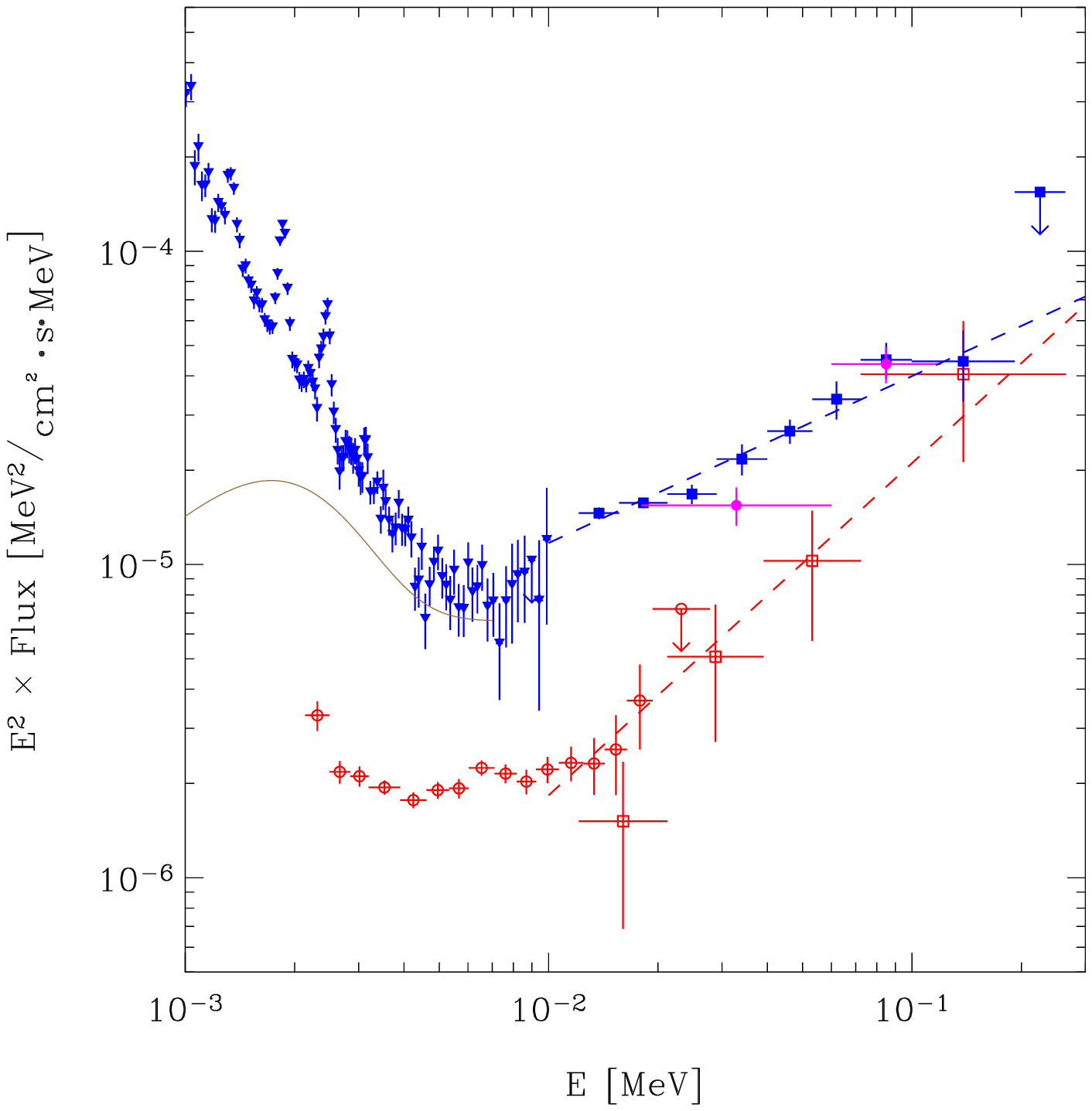}}
  \caption{\label{high_energy_spc}A $\nu F_{\nu}$ spectral representation of the total pulsed high-energy emission
  from \axp\ is shown in red (RXTE PCA; open circles, RXTE HEXTE; open squares). The total spectrum from 
  the Kes73/\axp\ complex is represented in blue (triangles; XMM EPIC PN, filled squares; RXTE HEXTE). The total 
  (pulsed plus DC) 1-7 keV X-ray spectrum from \axp\ \citep{morii03} is plotted as a solid dark orange line. The two magenta 
  flux points are INTEGRAL IBIS ISGRI measurements given in \citet{molkov04}. Fits ($>10$ keV) to the total complex (blue) and 
  total pulsed (red) spectra of \axp\ are drawn as dashed lines.}
\end{figure}

\section{Spectral analysis}

In this section high-energy ($\sim 2-300$ keV) spectra are presented 
for: 1) the total pulsed emission from \axp\ based on RXTE PCA and HEXTE data, and 2) 
the total high-energy emission from the Kes 73 supernova remnant plus 
the embedded AXP \axp\ using RXTE HEXTE on/off data and XMM Newton PN data. 


\subsection{Pulsed X-ray spectrum of \axp}
The number of pulsed counts in the differential PCA/HEXTE energy bands has been determined by fitting a truncated Fourier series, 
$a_0+\sum_{k=1}^{N} a_k \cos(2\pi k \phi) + b_k \sin(2\pi k \phi) $, with $\phi$ the pulse phase, to the measured pulse phase distributions. It turned out that 3 
harmonics ($N=3$) were sufficient to describe the measured distributions accurately for all energy intervals. For example, in Figs. \ref{pca_pulse_template} and 
\ref{hexte_pulse_prof_gt50} the best fit truncated Fourier series are shown as dashed lines for the PCA 3.9-7.8 keV and HEXTE $> 50.3$ keV pulse phase 
distributions, respectively.

In case of the PCA we derived for each PCU the energy response matrix (energy redistribution including the  sensitive area) for the combination of observations
listed in Table  \ref{obs_table} and subsequently took the different PCU (screened) exposure times into account in the construction of the weighted PCU-combined 
energy response. The pulsed excess counts per energy band are fitted in a procedure assuming a power-law  photon spectrum folded through the PCU-combined 
energy response. The best fit power-law model to the PCA pulsed data  has a photon index of $1.93\pm 0.01$. The derived 2-10 keV total pulsed 
flux\footnote{Spectral cross-calibration between the BeppoSAX MECS and RXTE PCA instruments indicate that the normalization of the PCA spectra is about 
20\% higher than that of the MECS. The PCA fluxes are therefore multiplied by 0.81 \citep[see e.g.] [about this subject]{kuiper04}} is 
($5.20\pm 0.14)\cdot 10^{-12}$ erg/cm$^2$s, which is consistent with the 2-10 keV pulsed 
flux of ($5.70\pm 0.70)\cdot 10^{-12}$ erg/cm$^2$s given in \citet{gotthelf02} who used a smaller dataset and a different extraction method (on/off).

In the spectral ``deconvolution" process of the HEXTE total pulsed counts the on-axis\footnote{It was not necessary to derive off-source HEXTE response matrices 
taking into account the reduction in the sensitive area due to the collimator response because all observations were performed nearly on-axis.} cluster A and B 
energy response matrices have been employed taking into account the (slightly) different screened on-source exposure times of each cluster. The exposure 
times have been corrected for the considerable deadtime effects. 

The total pulsed PCA and HEXTE flux measurements are shown  in Fig. \ref{high_energy_spc} as red open circles and squares, respectively. In this $\nu F_{\nu}$
spectral representation the hardening of the pulsed flux near 10 keV is striking. A power-law model fitted  to the PCA/HEXTE total pulsed flux measurements for energies above 
10 keV yielded a photon index of $0.94\pm 0.16$. Such a spectral hardening has not been observed yet for any of the AXPs currently known. 


\subsection{Total X-ray spectrum of Kes 73 plus \axp}
For HEXTE a model independent way for deriving the total flux from the Kes 73/\axp\ complex exists based on the source on/off observations.
The off-source observations make it possible to obtain instantaneous background estimates. Subtracting the properly scaled background 
contribution (every 32 s on-source observation is accompanied by a 28 s off-source observation; 4 s is consumed in the slew process) from 
the on-source measurements yielded the total excess counts from the Kes 73/\axp\ complex for energies above $\gtap 10$ keV.
These excess counts have been converted to flux values using similar procedures as discussed before and the resulting flux measurements
are shown as blue filled squares in Fig. \ref{high_energy_spc}. A power-law fit through these points yielded a photon index of $1.47\pm 0.05$,
slightly softer than the pulsed emission for energies above 10 keV.

In order to extend the total emission spectrum towards lower energies we analysed archival XMM Newton data from an observation 
targetted at \axp. We used data taken on October 5, 2002, starting at 03:16:47 UTC. We processed the data using the latest version 
of the XMM-Newton analysis software, SAS 6.0.0. For the analysis, we first produced light curves in different energy bands
of a region in the sky far from the supernova remnant to check for solar flares that might have affected the observation. Since no flares
were detected in the light curves, we used the total length of the observation, $\sim 4000$ s of on source time, for the rest of the
analysis. We produced images and spectra both for the EPIC PN and the two EPIC MOS instruments, and checked that all instruments 
yielded consistent results. Since the EPIC PN has the largest overall effective area and is sensitive to higher energies than MOS, we 
only concentrated on this instrument for the rest of our analysis.

We extracted a spectrum of a circular region of 2 arcmin radius centered on \axp\ containing both the AXP and the supernova
remnant. We checked that the spectrum is not affected by pile up\footnote{Pile up occurs when two or more photons hit the same CCD
pixel within a read-out cycle. In that case the instruments count those photons as single event with an energy approximately equal to the sum
of the energies of the individual photons. Pile up reduces the apparent flux of the source and at the same time makes the spectrum appear
harder than it really is.}. To produce a background spectrum we selected an area near the supernova remnant, but sufficiently far from it to
avoid contamination from the supernova itself.
Next we produced an energy  response matrix and an ancilliary response file, the latter properly weighted using the surface brightness
distribution of the extended emission of the source over the detector. We rebinned the spectrum such that we had about 3 bins per 
resolution element. Since the source is relatively bright, with this rebinning we obtained at least 25 counts per bin in the range 0.3 -- 7 keV, 
and at least 16 counts per bin up to 11 keV.
The ``deconvolved'' spectrum ($\nu F_{\nu}$) is shown in Fig. \ref{high_energy_spc}:  it displays a relatively bright continuum, on top of which 
several emission lines of Mg, Ar, S, and Si coming from the supernova remnant can be seen.


\section{Discussion}

We have presented the surprising discovery of very hard, photon index 0.94 $\pm$ 0.16, 
pulsed hard X-ray emission extending up to energies of $\sim$ 150 keV from the AXP \axp,
obtained using archival monitoring observations by the PCA and HEXTE aboard RXTE. Such hard, non-thermal
emission can only originate within the magnetosphere of a neutron star/magnetar. In fact, the total pulsed X-ray spectrum
of \axp\ is very reminiscent to that of young spin-down powered pulsars. Particularly the spectrum of the Vela pulsar
with a blackbody spectrum at lower X-ray energies and a hard power-law extension to higher energies looks very similar.
These hard power-law  spectral extensions with photon indices  of $\sim 1$ are also found for the young (radio weak/quiet) 
spin-down powered  pulsars PSR B1509-58, PSR B1846-0258, PSR J1811-1925, PSR J1930+1852, all associated with SNRs
, like  \axp, and the old millisecond pulsars PSR J0218+4232, PSR B1821-24 and PSR B1937+21.

Such hard X-ray emission from AXPs has hardly been considered in the modelling of AXPs. However, \citet{cheng01}
studied the production of high-energy gamma-ray radiation in the outer magnetospheres of AXPs. They argue that due to the
strong field of a magnetar, the gamma-ray emission rooted at the polar caps will be quenched. However, far away from the 
pulsar surface, i.e. in the outer gaps, the gamma-rays could be emitted because the local magnetic field will drop 
below the critical quantum limit. The production process would be curvature radiation by the acceleration of electrons/positrons, 
resulting from the collisions between high-energy photons from the outer gap and the soft X-rays originating from the stellar surface. 
They applied their model calculations also to the case of \axp, and predict an integral gamma-ray photon flux for energies 
above 100 MeV of $\sim$ 8.7 $\times$ 10$^{-9}$ (d/6 kpc)$^{-2}$ cm$^{-2}$ s$^{-1}$, which is below the EGRET sensitivity 
but above the sensitivity of GLAST. The very-hard pulsed spectrum we measured with HEXTE might be consistent with this interpretation.

We studied the energy dependence of the pulse profile morphology and confirm the findings by \citet{morii03}, namely
the pulse profile is double-peaked, and the second pulse has a harder spectrum for energies below $\sim$ 7 keV. However,
this trend does not continue above $\sim 10$ keV. Within the statistics, both pulses remain visible at about the same
strengths up to $\sim$ 150 keV. Significant variations with energy of pulse profile morphologies below $\sim 8$ keV 
have been reported earlier for three other AXPs (4U 0142+61, 1RXS 1708-4009 and 1E 2259.1+586) by \citet{gavriil02a}.

The set of spectra shown in Fig. \ref{high_energy_spc} summarizes our present knowledge on the contributions from
the different components constituting the total emission from the direction of Kes 73 and AXP \axp, i.e.
the pulsed and DC components of \axp, and the emission from SNR Kes 73.
In addition to the total pulsed spectrum (PCA/HEXTE) and the pulsed+DC+Kes-73 spectrum (XMM EPIC PN/HEXTE) derived in this work,
we show the INTEGRAL IBIS ISGRI flux values for energies 18--120 keV \citep{molkov04}, and the Chandra ACIS spectrum \citep{morii03}
for the pulsed+DC spectrum of \axp. We note that the INTEGRAL flux points are consistent with our total HEXTE spectrum
(one should realize that these preliminary INTEGRAL IBIS ISGRI flux values have still systematic uncertainties of $\sim$ 10\%).

For energies up to about 7 keV we have information on the relative contributions:  
1) the pulsed fraction of \axp\ is $\sim$ 25\%, 2) there is very little 
 margin for emission from SNR Kes 73 around 7 keV. This suggests that the HEXTE spectrum above 10 keV is dominated
 by emission from \axp\ (pulsed+DC), and above $\sim 100$  keV the spectrum is totally due to pulsed emission.
 This would mean that also the DC emission from \axp\ extends up to $\sim 100$  keV!
 An alternative interpretation is that the relative contribution from Kes 73 increases again above 10 keV due to a new hard component.
 Although the latter seems unlikely, higher spatial-resolution measurements at energies above 10 keV would be required
 to exclude this possibility.

 If we assume that the contribution from Kes 73 is small or negligible above 10 keV, the challenge is now to explain
 the hard DC emission from \axp. Most plausible might be an origin in the magnetosphere, namely for
 the situation that the pulsed emission is ``on" for all phases. In the competing polar cap \citep[e.g. as discussed for AXPs by][] 
 {zhang00} and outer gap \citep[see e.g.][]{cheng01} scenarios explaining the high-energy emission from spin-down powered pulsars 
 there are geometries  for which this is indeed possible. Allowing for a small contribution from Kes 73, e.g. up to about 30 keV, and 
 knowing  that pulsar spectra depend on pulse phase, detailed model calculations are required to investigate this interpretation further.


\acknowledgments
This research has made use of data obtained from the High Energy Astrophysics Science Archive Research Center (HEASARC), 
provided by NASA's Goddard Space Flight Center.


\clearpage 
\end{document}